\documentclass[journal]{IEEEtran}

\usepackage{latexsym}
\usepackage{graphicx}
\usepackage{amsfonts,amssymb,amsmath}
\usepackage{ctable} 
\usepackage{mathtools, cuted}
\usepackage{hyperref}
\usepackage[T1]{fontenc}
\usepackage{cite}
\usepackage{subcaption}
\usepackage{comment}
\usepackage{diagbox}

\usepackage{amsthm}
\usepackage{overpic}
\usepackage{steinmetz}
\usepackage{array}
\usepackage{url}
\usepackage{color}

\usepackage{algorithm}
\usepackage{algpseudocode}
\algnewcommand{\LongState}[1]{\State \parbox[t]{0.88\linewidth}{#1}}

\usepackage{xcolor}
\usepackage{soul}

\usepackage{linegoal}

\theoremstyle{plain}

\def\Htran{\mbox{\tiny $\mathrm{H}$}}
\def\Ttran{\mbox{\tiny $\mathrm{T}$}}

\algrenewcommand\algorithmicrequire{\textbf{Input:}}
\algrenewcommand\algorithmicensure{\textbf{Output:}}

\IEEEoverridecommandlockouts

\begin{document}
\bstctlcite{BSTcontrol}
\title{Non-Uniform Array  Configuration and Antenna Deactivation for Energy-Efficient Sensing}
\author{Parisa Ramezani, Meesam Jafri, Marina Petrova, \textit{Member, IEEE,}  and Emil Bj\"ornson, \textit{Fellow, IEEE} 
\thanks{ Parisa Ramezani is with the School of Electrical, Mechanical and Biomedical Engineering, University of Technology Sydney, 2007 Sydney, Australia (email: parisa.ramezani@uts.edu.au). Meesam Jafri and Emil Bj\"ornson are with the Department
of Communication Systems, KTH Royal Institute of Technology, 100 44
Stockholm, Sweden (email: meesam@kth.se, emilbjo@kth.se). Marina Petrova is with KTH and the Chair of Mobile Communications and Computing, RWTH Aachen University, 52056 Aachen, Germany (email: petrovam@kth.se, petrova@mcc.rwth-aachen.de).
}
}

\maketitle
\begin{abstract}
This paper proposes a non-uniform array design framework for energy-efficient multi-target sensing by jointly optimizing the antenna configuration and the transmit covariance matrix (TCM). To begin with, a limited number of antennas are selected from a set of antenna locations to maximize the minimum mainlobe-to-peak-sidelobe-level ratio (MMPSR) over a sensing region. Next, the antenna configuration is optimized using a simulated annealing (SA)-based search, while a TCM optimization problem is formulated using Charnes-Cooper transformation and solved as a convex semidefinite program for each fixed configuration. Subsequently, an adaptive antenna deactivation strategy is developed to determine the maximum number of antennas that can be switched off while satisfying an MMPSR requirement. Finally, simulation results demonstrate that the proposed non-uniform array design outperforms fixed uniform and non-uniform array benchmarks and requires fewer active antennas to achieve a required sensing performance, thereby supporting energy-efficient sensing.
\end{abstract}

\vspace{-10pt}
\section{Introduction}
Wireless systems are increasingly employing large antenna arrays as reflected in the emergence of ultra-massive multiple-input multiple-output (MIMO) \cite{Yang2024Signal}, extremely large-scale MIMO \cite{Lu2022NearField}, and gigantic MIMO architectures \cite{Emil2025Enabling}. The increased number of antenna elements provides additional spatial degrees of freedom that can enhance sensing performance in terms of target detection probability, sidelobe suppression, and spatial resolution. However, increasing the number of antennas also increases hardware cost, power consumption, and signal processing complexity. 

Recent works have investigated reconfigurable array architectures, such as fluid and movable antenna systems, where a finite number of antennas can be repositioned within a spatial region \cite{Zhang2025Successive,Han2025Cell,Li2026Low, Zhu2023Movable, Ning2025Movable}. These architectures provide additional freedom in controlling the array geometry; however, they require antenna repositioning mechanisms and may incur movement latency, control overhead, and additional power consumption.

A practical alternative is to place antennas non-uniformly at a finite set of fixed locations and electronically activate a selected subset, enabling a trade-off between sensing performance and energy consumption. The study of such discrete array architectures dates back several decades.
The works in \cite{Skolnik1964Planar}
and \cite{Murino1996Synthesis} select a fixed number of antenna elements from a finite set of available positions to obtain non-uniform arrays with reduced sidelobe levels. Owing to their practical relevance, the design of such discrete array architectures has recently regained considerable attention. The work in \cite{Palaiologos2024Robust} jointly designs non-uniform transmit and receive arrays for line-of-sight MIMO systems by selecting antenna locations from finite antenna grids of arbitrary geometry. The objective therein is to maximize the minimum channel capacity over a range of transmit distances. In \cite{Wang2026Harnessing}, a flexible non-uniform array design is considered for monostatic integrated sensing and communication (ISAC) systems, where each antenna can operate in transmit, receive, or inactive mode. The antenna-mode assignment and ISAC beamforming are jointly optimized to maximize the communication sum rate subject to sensing, power, and antenna-mode constraints.

In this paper, we consider a multi-target sensing system and adopt the minimum mainlobe-to-peak-sidelobe-level ratio (MMPSR) as the sensing metric. The MMPSR maximizes the weakest target response relative to the peak sidelobe level. In contrast to conventional peak-sidelobe minimization \cite{Skolnik1964Planar,Murino1996Synthesis}, MMPSR is better suited for multi-target sensing scenarios as it avoids configurations that suppress sidelobes well but provide insufficient illumination to one or more targets.
We propose a simulated annealing (SA)-based framework for the joint optimization of a non-uniform antenna configuration and the transmit covariance matrix (TCM). Building on the antenna position search in \cite{Murino1996Synthesis}, our framework additionally optimizes the TCM for each configuration and extends the design to arbitrarily shaped array geometries and general sensing regions. SA is particularly attractive when only a small fraction of the antenna locations is activated, since each trial optimizes a reduced TCM whose dimension is determined by the number of active antennas, rather than by the total number of locations. In contrast, binary relaxation-based approaches, such as the one presented in \cite{Wang2026Harnessing}, involve optimization variables whose dimensions scale with the total number of antenna locations, potentially leading to high computational complexity when the set of antenna locations is large.
To further reduce energy consumption, we develop a dynamic programming-based adaptive antenna deactivation strategy. Motivated by the predecessor-based array design principle in \cite{Palaiologos2024Robust}, the proposed method progressively constructs candidate deactivation sequences and stores their MMPSR values and predecessor information in offline lookup matrices, which are then used to determine the maximum number of antennas that can be deactivated for a given MMPSR requirement. Numerical results demonstrate that the proposed non-uniform array design achieves higher MMPSR than fixed uniform and non-uniform arrays. The results further illustrate that the proposed deactivation strategy can substantially reduce the number of active antennas required to meet an MMPSR requirement.
\vspace{-5pt}
\section{System Model and Problem Formulation}
We consider a multi-target sensing system with a transmitter equipped with an arbitrarily shaped non-uniform array, whose active antennas are selected from predefined locations to illuminate a sensing region using probing waveforms.
Let $\mathcal P
    =
    \left\{
    \mathbf p_{1},
    \mathbf p_{2},
    \ldots,
    \mathbf p_{N}
    \right\}$
denote the set of antenna locations, where $N$ is the number of  locations on the antenna array and $\mathbf p_{n} = [ x_{n},  y_{n},  z_{n}]^{\Ttran}\in \mathbb R^3$. 
Furthermore $\mathbf q = [x_q,y_q,z_q]^{\Ttran} \in \mathbb R^3$ denotes an arbitrary point in the sensing region. The array response vector for a point $\mathbf q$ is given by 
    $\mathbf a(\mathbf q)
    =
    \left[
    e^{-j\frac{2\pi}{\lambda}d_{1}(\mathbf q)},
    \ldots,
    e^{-j\frac{2\pi}{\lambda}d_{N}(\mathbf q)}
    \right]^{\Ttran}
    \in \mathbb C^{N},$
where $\lambda$ is the carrier wavelength and  $d_{n}(\mathbf q) = \|\mathbf q - \mathbf p_{ n}\|$ is the distance between the $n$-th  antenna and point $\mathbf q$. 

To model the activation of a finite number of antennas over the available locations, antenna location $n$ is associated with a binary selection variable $s_n$, where $s_n = 1$ indicates that the $n$th antenna is activated and $s_n = 0$ indicates deactivation. Due to hardware cost, power consumption, and signal processing complexity, the number of active antennas is restricted to $M$, with $M \ll N$.
Let $\mathbf x \in\mathbb C^{N}$ denote the probing signal with $\mathbf Q = \mathbb E[\mathbf x \mathbf x^{\Htran}] \in\mathbb C^{N \times N}$ as its TCM.
The objective is to jointly design the transmit array configuration and the TCM to maximize the sensing MMPSR by enhancing the minimum beampattern response over the target locations while suppressing the maximum response over sidelobe regions.

The sensing response of a target located at point $\mathbf q$ is characterized by the transmit beampattern generated by the selected antenna locations and the TCM. The sensing beampattern is defined as $P(\mathbf q) =    \mathbf a^{\Htran}(\mathbf q)\mathbf Q\mathbf a(\mathbf q)$, which can be equivalently written as $\operatorname{tr}     \left( \mathbf A(\mathbf q)\mathbf Q  \right)$, where $\mathbf A(\mathbf q)=\mathbf a(\mathbf q)\mathbf a^{\Htran}(\mathbf q)$.
Let the set of $K$ target locations be denoted by $\mathcal T
    =
    \left\{
    \mathbf q_1^{\rm tar},
    \mathbf q_2^{\rm tar},
    \ldots,
    \mathbf q_K^{\rm tar}
    \right\}.$
The sidelobe region is obtained by excluding the target locations and their mainlobe guard regions from the sensing region, and is discretized into a set of $L$  sampling points as
$
    \mathcal S
    =
    \left\{
    \mathbf q_1^{\rm sl},
    \mathbf q_2^{\rm sl},
    \ldots,
    \mathbf q_L^{\rm sl}
    \right\}.$
Consequently, the sensing MMPSR is defined as
\begin{equation}
\label{eq:MMPSR}
    \mathrm{MMPSR}
    =
    \frac{
    \min\limits_{k=1,\ldots,K}
    P(\mathbf q_k^{\rm tar})
    }{
    \max\limits_{\ell=1,\ldots,L}
    P(\mathbf q_\ell^{\rm sl})
    + \delta},
\end{equation}
where $\delta > 0$ prevents the denominator from becoming zero. The MMPSR maximization problem is formulated as
\begin{subequations}
\label{eq:PSR_max}
\begin{align}
\max_{\mathbf Q,\mathbf s,t,z}
~&
\frac{t}{z+\delta}
\label{eq:prob_obj}
\\
\mathrm{s.t.}~
&
\operatorname{tr}
\left(
\mathbf A(\mathbf q_k^{\rm tar})\mathbf Q
\right)
\geq t,
~\forall k,
\label{eq:prob_tar}
\\
&
\operatorname{tr}
\left(
\mathbf A(\mathbf q_\ell^{\rm sl})\mathbf Q
\right)
\leq z,
~\forall \ell,
\label{eq:prob_sl}
\\
&
\mathbf Q\succeq \mathbf 0,
\label{eq:prob_psd}
\\
&
[\mathbf Q]_{n,n}
\leq
P_{\rm ant}^{\max}s_n,
~n=1,\ldots,N,
\label{eq:prob_selection}
\\
&
\sum_{n=1}^{N}s_n=M,
\label{eq:prob_num}
\\
&
s_n\in\{0,1\},
~n=1,\ldots,N,
\label{eq:prob_binary}
\end{align}
\end{subequations}
where $\mathbf{s}=\left[s_{1}, s_{2}, \cdots, s_{N}\right]^{\Ttran}$ denotes the antenna-selection vector. The scalar quantities $t$ and $z$ are auxiliary variables that lower-bound the minimum target response and upper-bound the peak sidelobe response, respectively. Hence, maximizing the objective in \eqref{eq:prob_obj} is equivalent to maximizing the MMPSR in \eqref{eq:MMPSR}. Constraint \eqref{eq:prob_selection} couples the antenna-selection variables with the TCM, where it forces inactive locations to radiate zero power and limits the power of each active antenna. Furthermore \eqref{eq:prob_num}--\eqref{eq:prob_binary} impose the antenna configuration constraints.
In the next section, we address this mixed-integer problem by separating the TCM design from the antenna configuration design. For each fixed antenna configuration, the TCM is optimized to maximize the corresponding MMPSR, while SA is employed to iteratively update the active antenna locations.
\section{Antenna Configuration and TCM Design} 

For a given antenna configuration, the problem \eqref{eq:PSR_max} reduces to a linear fractional semidefinite program (SDP) with respect to $\mathbf{Q}$. Since the inactive rows and columns of $\mathbf Q$ are zero, the optimization problem can be equivalently expressed in terms of the reduced-dimension TCM, $\widetilde{\mathbf Q}\in\mathbb C^{M\times M}$, corresponding to the active antenna locations. Let $\widetilde{\mathbf a}(\mathbf q)\in\mathbb C^M$ denote the corresponding reduced array response vector and define
$\widetilde{\mathbf A}(\mathbf q)=
\widetilde{\mathbf a}(\mathbf q)
\widetilde{\mathbf a}^{\Htran}(\mathbf q)$. The resulting covariance-design problem is a linear-fractional SDP. 
Since $z+\delta>0$, the Charnes--Cooper transformation can be applied to reformulate the linear-fractional TCM design problem as an equivalent convex SDP. This requires only a single convex optimization solve per antenna configuration and avoids a bisection-based solution, which would require solving a sequence of TCM optimization problems for each configuration. 
Introducing the Charnes--Cooper scaling variable $\xi$ and the transformed variables as
\begin{equation}
\xi=\frac{1}{z+\delta},
\quad
\overline{\mathbf Q}=\xi\widetilde{\mathbf Q},
\quad
\bar t=\xi t,
\label{eq:CC_variables}
\end{equation}
and using $\xi(z+\delta)=1$, the TCM optimization problem can be equivalently written as
\begin{subequations}
\label{eq:CC_problem}
\begin{align}
\max_{\overline{\mathbf Q},\bar t,\xi}
~&
\bar t
\\
\mathrm{s.t.}~
&
\operatorname{tr}
\left(
\widetilde{\mathbf A}(\mathbf q_k^{\rm tar})
\overline{\mathbf Q}
\right)
\geq \bar t,
~\forall k,
\\
&
\operatorname{tr}
\left(
\widetilde{\mathbf A}(\mathbf q_\ell^{\rm sl})
\overline{\mathbf Q}
\right)
\leq
1-\delta\xi,
~\forall \ell,
\\
&
\overline{\mathbf Q}\succeq\mathbf 0,
\\
&
[\overline{\mathbf Q}]_{m,m}
\leq
\xi P_{\rm ant}^{\max},
~m=1,\ldots,M,
\\
&
\xi\geq0.
\end{align}
\end{subequations}
Problem \eqref{eq:CC_problem} is a convex SDP and can be solved using standard interior-point-based solvers. The optimized TCM and MMPSR are finally recovered as $\widetilde{\mathbf Q}^\star =\frac{\overline{\mathbf Q}^{\star}}{\xi^\star}$ and $\gamma^\star = \bar t^\star$, respectively.

To determine the antenna locations, we employ an SA-based approach \cite{Rutenbar1989Simulated}. Starting from a random configuration with $M$ active antennas, new configurations are generated by swapping active and inactive locations, with the TCM optimized via \eqref{eq:CC_problem} for each configuration.
Improving configurations are always accepted, while non-improving ones are accepted with a temperature-dependent probability, reducing the likelihood of convergence to poor local optima. A geometric cooling schedule is used, with $T_i=\alpha^{i-1}T_0$, $i=1,\ldots,I_{\max}$, or equivalently $T_{i+1}=\alpha T_i$, where $T_0$ is the initial temperature and $0<\alpha<1$ is the cooling factor. At iteration $i$, after generating a trial configuration and solving the corresponding covariance-design problem, the MMPSR variation is $\Delta\gamma=\gamma^\star-\gamma^{\rm cur}$, where $\gamma^\star$ and $\gamma^{\rm cur}$ are the optimized MMPSRs of the trial and current configurations, respectively.
 The acceptance probability of the new configuration is defined as
\begin{equation}
p_{\rm acc}^{(i)}
=
\begin{cases}
1, & \Delta\gamma\geq 0,\\[1mm]
\exp\!\left(\dfrac{\Delta\gamma}{T_i}\right),
& \Delta\gamma<0.
\end{cases}
\end{equation}

We employ a hybrid swap strategy with larger changes in early iterations and finer refinements later. Let $\mathcal N^{\rm cur}$ denote the current active-antenna set and $\mathcal N^{\star}$ the trial set at iteration $i$. The trial set is generated as $\mathcal N^{\star}=(\mathcal N^{\rm cur}\setminus\mathcal U_i)\cup\mathcal V_i$, where $\mathcal U_i\subseteq\mathcal N^{\rm cur}$ and $\mathcal V_i\subseteq\{1,\ldots,N\}\setminus\mathcal N^{\rm cur}$ are the removed and activated index sets, respectively, with $|\mathcal U_i|=|\mathcal V_i|=c_i$. We set $c_i=2$ for $i\leq I_{\rm two}$ and $c_i=1$ otherwise, where $I_{\rm two}$ is the number of initial two-swap iterations. Thus, two-swap moves provide broader exploration initially, while one-swap moves enable finer local refinement as the temperature decreases. Algorithm~\ref{alg:SA_transmit} summarizes the proposed antenna configuration and TCM design. To reduce initialization sensitivity, multiple independent initializations can be used, retaining the solution with the highest final MMPSR.
\begin{algorithm}[t]
\footnotesize
\caption{\small Proposed Antenna Configuration and TCM Design}
\label{alg:SA_transmit}
\begin{algorithmic}[1]
\Require Candidate locations $\mathcal P$, target set $\mathcal T$,
sidelobe set $\mathcal S$, $P^{\max}_{\rm ant}$, $\delta$,
$I_{\max}$, $T_0$, $\alpha\in(0,1)$, and $I_{\rm two}$
\Ensure $\mathcal N^{\rm best}$, $\widetilde{\mathbf Q}^{\rm best}$,
$\gamma^{\rm best}$

\State Randomly generate $\mathcal N^{(0)}\subseteq\{1,\ldots,N\}$
with $|\mathcal N^{(0)}|=M$
\LongState{Solve the reduced covariance-design problem for
$\mathcal N^{(0)}$ to obtain $\widetilde{\mathbf Q}^{(0)}$
and $\gamma^{(0)}$}
\LongState{Set
$(\mathcal N^{\rm cur},\widetilde{\mathbf Q}^{\rm cur},
\gamma^{\rm cur})
\leftarrow
(\mathcal N^{(0)},\widetilde{\mathbf Q}^{(0)},\gamma^{(0)})$}
\LongState{Set
$(\mathcal N^{\rm best},\widetilde{\mathbf Q}^{\rm best},
\gamma^{\rm best})
\leftarrow
(\mathcal N^{\rm cur},\widetilde{\mathbf Q}^{\rm cur},
\gamma^{\rm cur})$ and $T\leftarrow T_0$}

\For{$i=1,\ldots,I_{\max}$}
    \State Set $c_i=2$ if $i\leq I_{\rm two}$, and $c_i=1$ otherwise
    \LongState{Randomly choose
    $\mathcal U_i\subseteq\mathcal N^{\rm cur}$ and
    $\mathcal V_i\subseteq\{1,\ldots,N\}\setminus\mathcal N^{\rm cur}$
    with $|\mathcal U_i|=|\mathcal V_i|=c_i$, and set
    $\mathcal N^\star=
    \mathcal N^{\rm cur}\setminus\mathcal U_i\cup\mathcal V_i$}
    \LongState{Solve the reduced covariance-design problem for
    $\mathcal N^\star$ to obtain $\widetilde{\mathbf Q}^\star$
    and $\gamma^\star$}
    \State Set $\Delta\gamma=\gamma^\star-\gamma^{\rm cur}$
    and draw $r\sim\mathcal U(0,1)$

    \If{$\Delta\gamma\geq0$ \textbf{or}
    $r\leq\exp(\Delta\gamma/T)$}
        \LongState{Set
        $(\mathcal N^{\rm cur},\widetilde{\mathbf Q}^{\rm cur},
        \gamma^{\rm cur})
        \leftarrow
        (\mathcal N^\star,\widetilde{\mathbf Q}^\star,\gamma^\star)$}
    \EndIf

    \If{$\gamma^{\rm cur}>\gamma^{\rm best}$}
        \LongState{Set
        $(\mathcal N^{\rm best},\widetilde{\mathbf Q}^{\rm best},
        \gamma^{\rm best})
        \leftarrow
        (\mathcal N^{\rm cur},\widetilde{\mathbf Q}^{\rm cur},
        \gamma^{\rm cur})$}
    \EndIf

    \State $T\leftarrow\alpha T$
\EndFor

\State \Return $\mathcal N^{\rm best}$,
$\widetilde{\mathbf Q}^{\rm best}$, $\gamma^{\rm best}$
\end{algorithmic}
\end{algorithm}
\vspace{-4mm}
\section{Adaptive Antenna Deactivation}
In this section, we develop an adaptive antenna deactivation framework that switches off antennas selected by Algorithm~\ref{alg:SA_transmit} while satisfying a specified MMPSR requirement. Let $\mathcal N^{\rm best}=\{n_1,\ldots,n_M\}$ denote the optimized antenna set and $\overline{\mathbf a}(\mathbf q)\in\mathbb C^M$ its corresponding array response vector. Each antenna $n_i$ is assigned a binary variable as
\begin{equation}
d_i=
\begin{cases}
1, & \text{if antenna $n_i$ is deactivated},\\
0, & \text{otherwise}.
\end{cases}
\end{equation}
Accordingly, $\sum_{i=1}^{M}d_i$ denotes the number of deactivated antennas.
For a specified MMPSR requirement $\gamma_{\rm req}$, the antenna deactivation
problem is formulated as
\begin{subequations}
\label{prob:antenna_deactivation}
\begin{align}
\underset{\mathbf Q,\mathbf d,t,z}{\operatorname{maximize}}
\quad&
\sum_{i=1}^{M}d_i
\label{prob:antenna_deactivation_obj}
\\
\text{subject to}\quad
&
\operatorname{tr}\!\left(
\overline{\mathbf A}(\mathbf q_k^{\rm tar})\mathbf Q
\right)
\geq t,
\quad \forall k,
\label{prob:antenna_deactivation_target}
\\
&
\operatorname{tr}\!\left(
\overline{\mathbf A}(\mathbf q_\ell^{\rm sl})\mathbf Q
\right)
\leq z,
\quad \forall \ell,
\label{prob:antenna_deactivation_sl}
\\
&
t\geq \gamma_{\rm req}(z+\delta),
\label{prob:antenna_deactivation_psr}
\\
&
\mathbf Q\succeq \mathbf 0,
\label{prob:antenna_deactivation_psd}
\\
&
[\mathbf Q]_{i,i}
\leq
P_{\rm ant}^{\max}(1-d_i),
\quad i=1,\ldots,M,
\label{prob:antenna_deactivation_power}
\\
&
d_i\in\{0,1\},
\quad i=1,\ldots,M,
\label{prob:antenna_deactivation_binary}
\end{align}
\end{subequations}
where $\overline{\mathbf A}(\mathbf q) = \overline{\mathbf a}(\mathbf q) \overline{\mathbf a}^{H}(\mathbf q)$. Constraint~\eqref{prob:antenna_deactivation_psr} guarantees that the achieved MMPSR is lower-bounded by $\gamma_{\rm req}$, while \eqref{prob:antenna_deactivation_power} forces the covariance entries corresponding to deactivated antenna to zero. 
Problem~\eqref{prob:antenna_deactivation} is a mixed-integer optimization problem. Obtaining its optimal solution requires an exhaustive evaluation of all possible antenna subsets which is computationally expensive. We propose a low-complexity dynamic programming-based framework, motivated by the array design approach in \cite{Palaiologos2024Robust}.
\begin{algorithm}[t]
\footnotesize
\caption{\small Construction of the Antenna Deactivation Lookup Matrices}
\label{alg:antenna_deactivation}
\begin{algorithmic}[1]
\Require Optimized antenna set
$\mathcal N^{\rm best}=\{n_1,\ldots,n_M\}$
\Ensure MMPSR matrix $\boldsymbol{\Gamma}$ and backtracking matrix $\mathbf B$

\State Initialize
$\boldsymbol{\Gamma}\in\mathbb R^{M\times(M-2)}$ with $-\infty$
and
$\mathbf B\in\mathbb N_0^{M\times(M-2)}$ with zeros

\For{$i=1,\ldots,M$}
    \LongState {Compute $\gamma(\{n_i\})$ by solving the covariance-design
    problem for $\mathcal N^{\rm best}\setminus\{n_i\}$}
    \State $[\boldsymbol{\Gamma}]_{i,1}
    \leftarrow \gamma(\{n_i\})$
    \State $[\mathbf B]_{i,1}\leftarrow 0$
\EndFor

\For{$c=2,\ldots,M-2$}
    \For{$i=1,\ldots,M$}
        \State
        $j^\star\leftarrow
        \underset{j:\,n_i\notin\mathcal R(j,c-1)}
        {\operatorname*{arg\,max}}\;
        \gamma\!\left(
        \mathcal R(j,c-1)\cup\{n_i\}
        \right)$
        \State
        $[\boldsymbol{\Gamma}]_{i,c}
        \leftarrow
        \gamma\!\left(
        \mathcal R(j^\star,c-1)\cup\{n_i\}
        \right)$
        \State
        $[\mathbf B]_{i,c}\leftarrow j^\star$
    \EndFor
\EndFor

\State \Return $\boldsymbol{\Gamma}$ and $\mathbf B$
\end{algorithmic}
\end{algorithm}

The proposed method first constructs two offline lookup matrices
$\boldsymbol{\Gamma}\in\mathbb R^{M\times(M-2)}
\quad\text{and}\quad
\mathbf B\in\mathbb N_0^{M\times(M-2)}$.
The matrices contain $M-2$ columns because at most $M-2$ antennas are deactivated, ensuring that at least two antennas remain active to preserve spatial beamforming capability. The column index $c$ represents the number of deactivated antennas, while each
row corresponds to one antenna in $\mathcal N^{\rm best}$. The entry
$[\boldsymbol{\Gamma}]_{i,c}$ stores the largest optimized MMPSR found when
$c$ antennas are deactivated and antenna $n_i$ is the last deactivated
antenna. The entry $[\mathbf B]_{i,c}$ stores the predecessor row in column
$c-1$ that produced this value.
For the first column, one antenna is deactivated at a time. Thus, for each $i=1,\ldots,M$, antenna $n_i$ is removed, the TCM is re-optimized for $\mathcal N^{\rm best}\setminus\{n_i\}$, and the resulting MMPSR is stored in $[\boldsymbol{\Gamma}]_{i,1}$. Since no predecessor exists for the first deactivation, $[\mathbf B]_{i,1}$ is set to zero.
For column $c \geq 2$, we build upon the solutions stored in column $c-1$. For a given row $i$, we assume that antenna $n_i$ is the $c$-th deactivated antenna. Then, we examine all feasible deactivation sets obtained from column $c-1$ that do not already contain $n_i$. For each such candidate, we deactivate antenna $n_i$ in addition to the previously deactivated antennas, re-optimize the TCM for the remaining active set, and compute the resulting MMPSR. The largest MMPSR among these candidates is stored in $[\boldsymbol \Gamma]_{i,c}$, while the row index of the corresponding entry in column $c-1$ is stored in $[\mathbf B]_{i,c}$. Therefore, $\boldsymbol \Gamma$ stores the best MMPSR values for different numbers of deactivated antennas, and $\mathbf B$ stores the information needed to recover the corresponding deactivation sequence by backtracking.

Algorithm~\ref{alg:antenna_deactivation} summarizes the proposed approach for constructing the lookup matrices. The method retains only the best predecessor associated with each row and deactivation level and is therefore a low-complexity approximation rather than an exhaustive solution of problem~\eqref{prob:antenna_deactivation}. In Algorithm~\ref{alg:antenna_deactivation}, $\mathcal R(j,c-1)$ denotes the set of $c-1$ deactivated antennas recovered by backtracking from row $j$ in
column $c-1$ of $\mathbf B$.
For any MMPSR requirement $\gamma_{\rm req}$, the number of
antennas to be deactivated is obtained from
\begin{equation}
c^\star
=
\max\left\{
c\in\{0,1,\ldots,M-2\}:
\widehat{\gamma}_c\geq\gamma_{\rm req}
\right\},
\end{equation}
where $\widehat{\gamma}_c=\max_i[\boldsymbol{\Gamma}]_{i,c}$. 
The corresponding deactivation set is recovered by backtracking through $\mathbf B$. The offline lookup matrices can then be used to determine the deactivation pattern for different MMPSR requirements without repeating the combinatorial search.
\vspace{-3mm}
\section{Numerical Results}
In this section, we evaluate the proposed schemes through numerical simulations. We first compare the SA-based non-uniform array design with four benchmarks: a compact uniform array with half-wavelength spacing around the same center, a large-aperture array where the active antennas are uniformly distributed over the full aperture, a fixed nested array constructed by combining dense and sparse array components according to the classical nested array design \cite{Pal2010nested,Pal2012nested}, and a random selection scheme that retains the best configuration among multiple trials.

\begin{figure}
    \centering
    \includegraphics[width=0.87\linewidth]{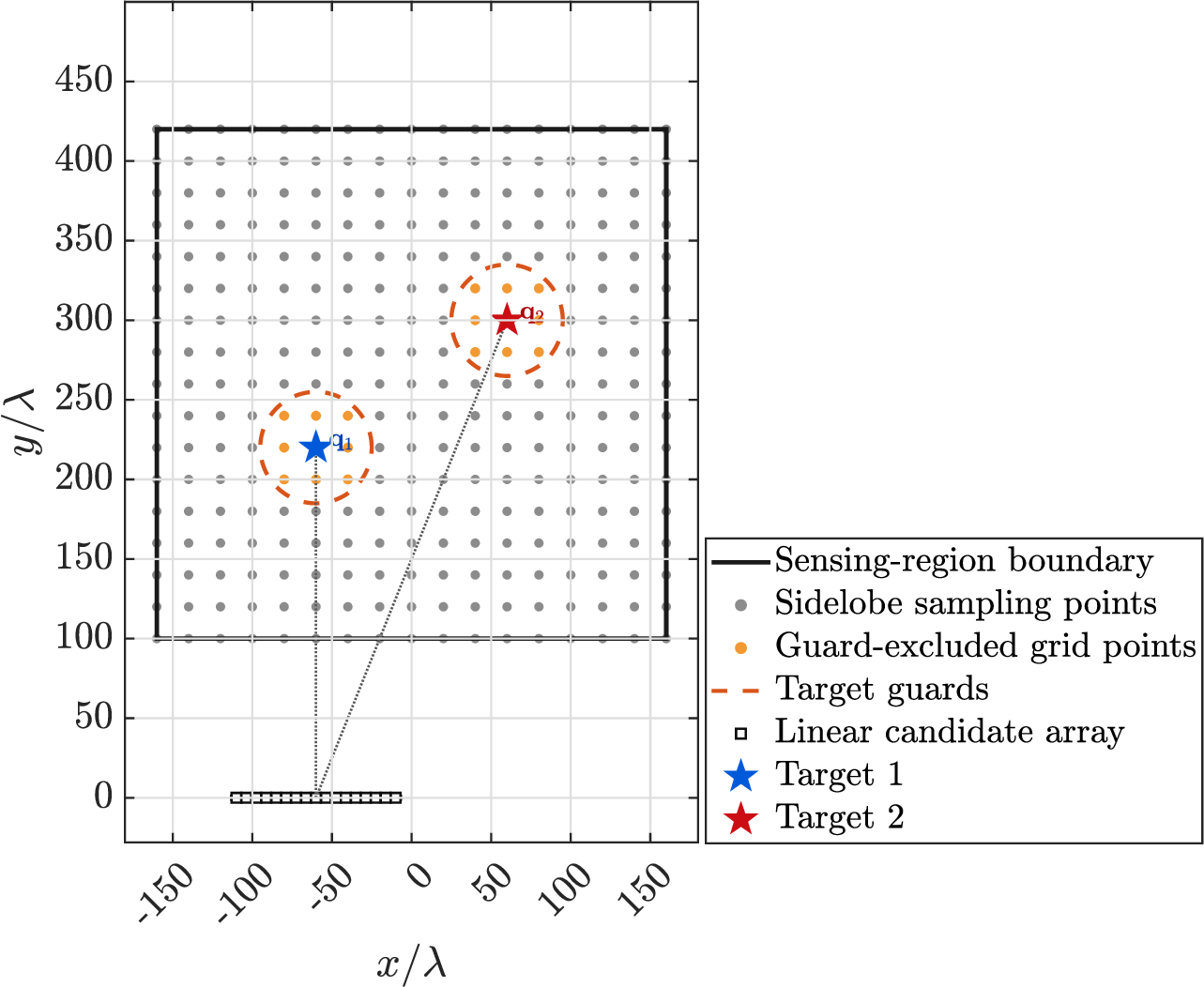}
    \caption{Simulation setup.}
\label{fig:sim_setup}
\vspace{-5mm}
\end{figure}

We consider a carrier frequency of $f_c=30\,$GHz. The simulation setup is depicted in Fig.~\ref{fig:sim_setup}. The array is centered at $(-60\lambda,0,0)$ and the spacing between antenna locations is $d = \lambda/2$. The sensing region is a two-dimensional area on the $x$-$y$ plane, defined by $-160\lambda\leq x\leq160\lambda$ and $100\lambda\leq y\leq420\lambda$, and is sampled with a grid spacing of $20\lambda$. Two target locations are considered at $(-60\lambda,220\lambda,0)$ and $(60\lambda,300\lambda,0)$. Around each target, a guard region with radius $35\lambda$ is excluded from the sidelobe sampling set.

We choose the initial temperature in Algorithm~\ref{alg:SA_transmit} such that a relative MMPSR degradation is accepted with a desired probability at the beginning of the algorithm. In particular, suppose that the MMPSR degradation of $\rho\gamma^{(0)}$ is accepted with probability $p_0$, where $\rho\in(0,1)$. Then, $\exp\!\left(-\frac{\rho\gamma^{(0)}}{T_0}\right)=p_0$
which yields $T_0=-\frac{\rho\gamma^{(0)}}{\ln p_0}$. We set $\rho = 0.1$ and $p_0 = 0.9$. Furthermore, the cooling factor is $\alpha = 0.95$. In all simulation setups, the number of SA iterations is set to $2000$. We set $P_{\rm ant}^{\max}=1\,$W, $\delta=10^{-4}$, and $I_{\rm two}=20$. For a fair comparison in terms of the number of evaluated antenna configurations, the random benchmark also evaluates $2000$ randomly generated configurations.

\begin{figure*}[!t]
    \centering
    \begin{subfigure}[t]{0.32\textwidth}
        \centering
        \includegraphics[width=\linewidth]{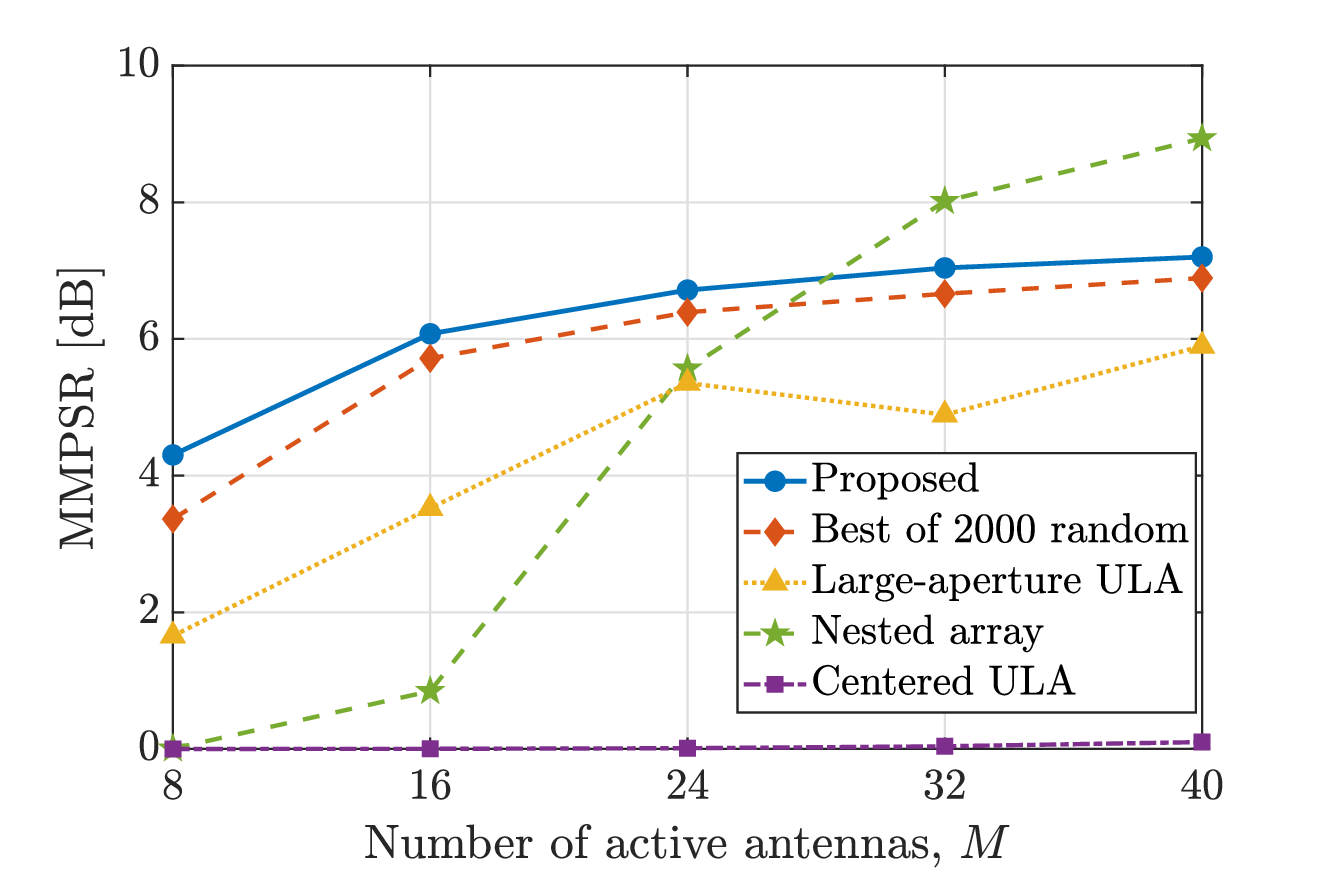}
        \caption{$N=128$ linear array.}
        \label{fig:LA_128}
    \end{subfigure}
    \hfill
    \begin{subfigure}[t]{0.32\textwidth}
        \centering
        \includegraphics[width=\linewidth]{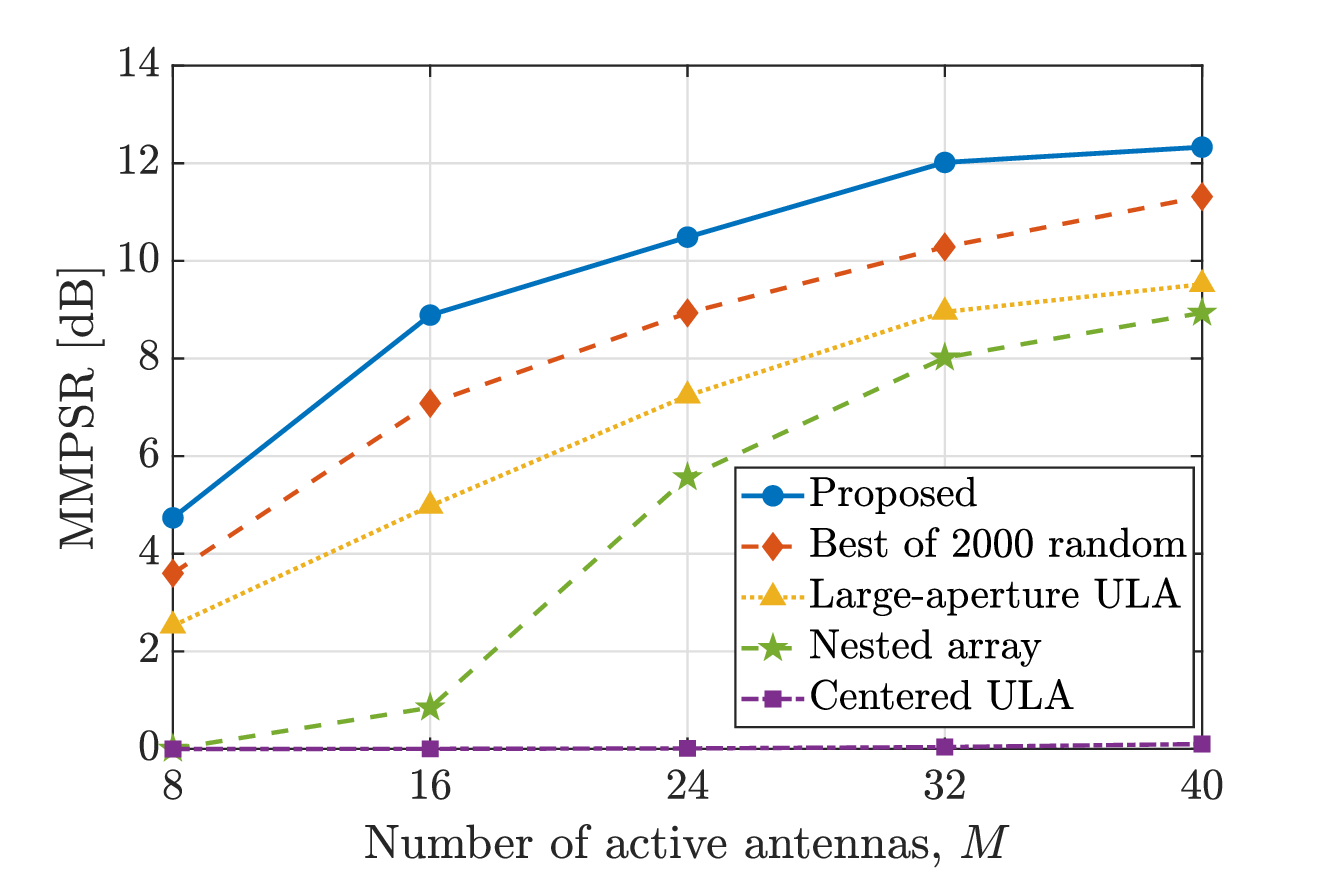}
        \caption{$N=256$ linear array.}
        \label{fig:LA_256}
    \end{subfigure}
    \hfill
    \begin{subfigure}[t]{0.32\textwidth}
        \centering
        \includegraphics[width=\linewidth]{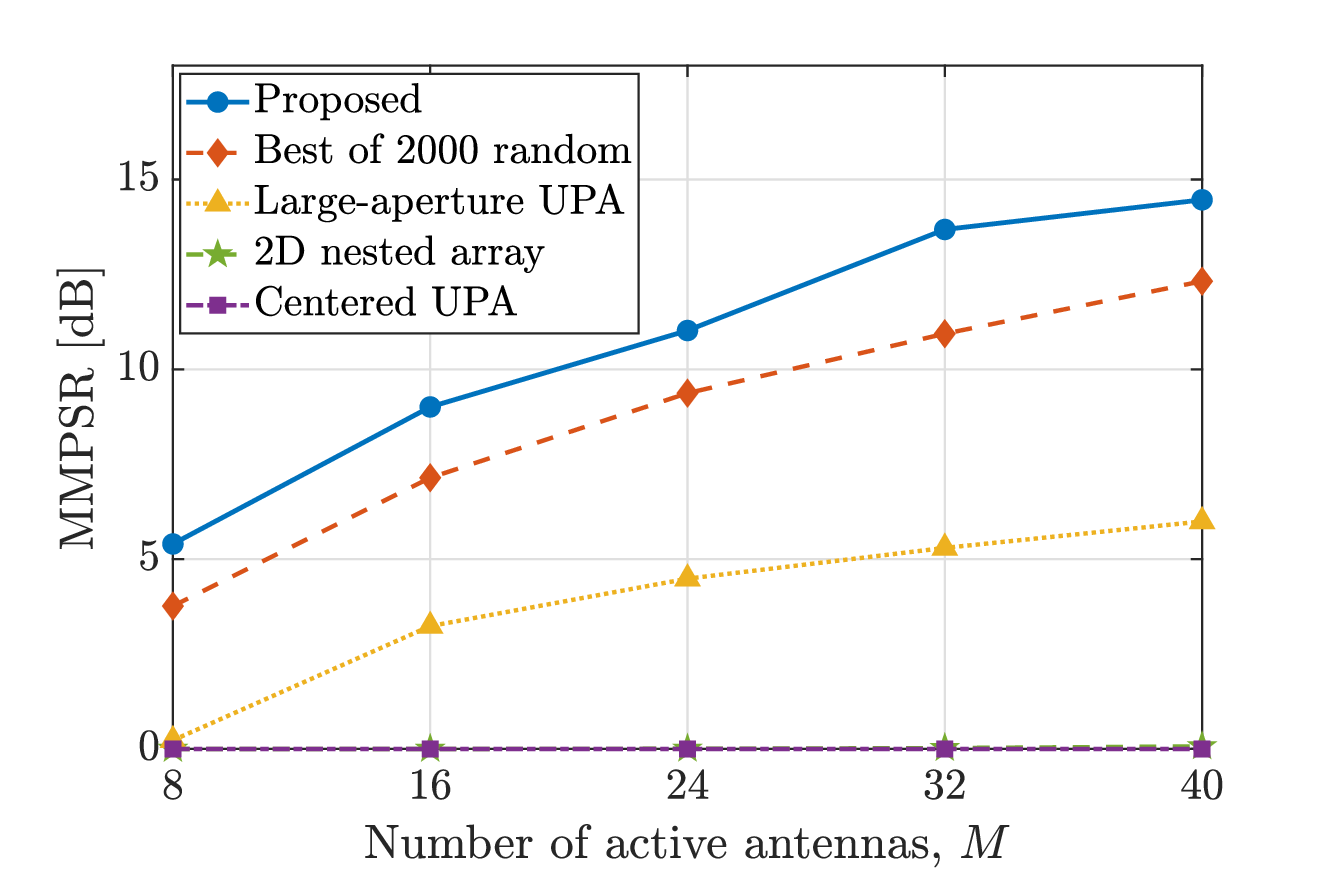}
        \caption{$181\times181$ planar array.}
        \label{fig:PA}
    \end{subfigure}
    \caption{MMPSR versus the number of active antennas.}
    \label{fig:PSR_all_arrays}
    \vspace{-4mm}
\end{figure*}
\begin{figure}
    \centering
    \includegraphics[width=0.75\linewidth]{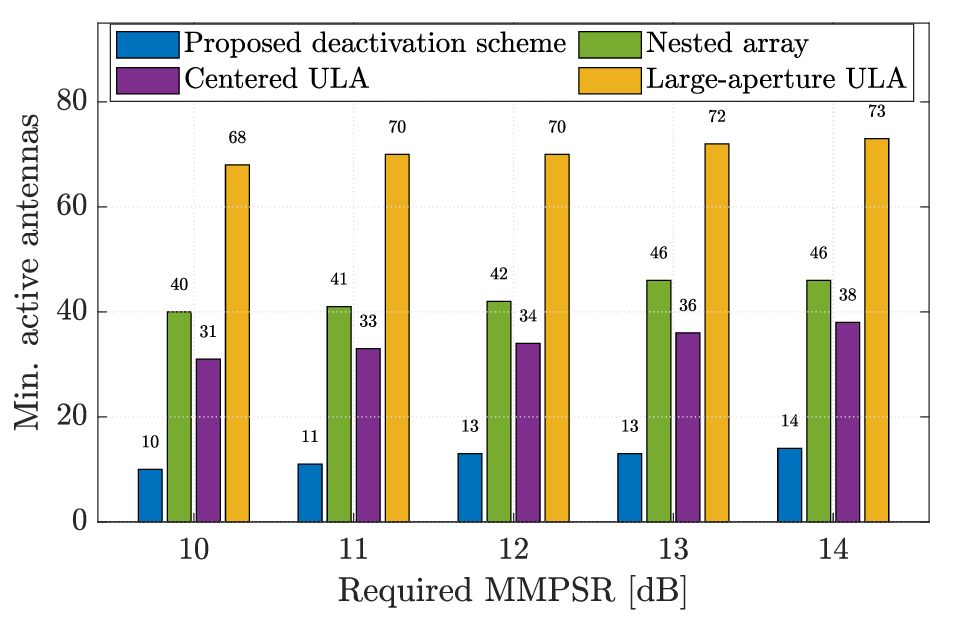}
   \caption{Minimum number of active antennas for different MMPSR values in the far-field scenario with $N=256$.}
\label{fig:min_active_ants}
\vspace{-5mm}
\end{figure}

Fig.~\ref{fig:PSR_all_arrays} compares the MMPSR versus the number of active antennas for the proposed and benchmark schemes. Fig.~\ref{fig:LA_128} considers $N=128$ antenna locations for the proposed, large-aperture ULA, and random-selection schemes, whereas the centered ULA and nested array have fixed geometries independent of $N$. The centered ULA performs poorly because its small aperture limits spatial resolution, while distributing antennas over the full aperture improves performance, although the resulting MMPSR can be non-monotonic in $M$ as changes in the discrete spacing pattern affect sidelobe interference unpredictably. The nested array improves rapidly with $M$ because its aperture grows approximately quadratically; at $M=32$ and $40$, its aperture is about $2.13$ and $3.30$ times that of the proposed array, respectively, explaining its higher MMPSR. At $M=24$, however, the proposed design performs better despite the nested array having a $1.22$-times larger aperture. The best-of-$2000$ random benchmark approaches the proposed scheme for some $M$ by retaining the best among many independently generated subsets, but does not exploit information across trials and depends on the sampled realizations.

Fig.~\ref{fig:LA_256} considers $N=256$ antenna locations with the same $M$ values. The larger aperture and greater selection flexibility improve the MMPSR relative to Fig.~\ref{fig:LA_128}, while the centered ULA and nested array remain unchanged because their geometries are independent of $N$. The proposed scheme outperforms the nested array for all considered $M$, despite the latter having a $1.64$-times larger aperture at $M=40$. The gap to the random benchmark also widens because the larger configuration space makes $2000$ random trials less likely to find a near-optimal solution.

Finally, Fig.~\ref{fig:PA} considers a \(181\times181\) planar grid whose diagonal aperture is approximately equal to the aperture of the \(N=256\) linear array. The planar array achieves a higher MMPSR due to greater spatial diversity and placement flexibility; at $M=40$, it attains $14.47$~dB versus $12.33$~dB for the $N=256$ linear array. The 2D nested array performs similarly to the centered ULA because of its relatively small aperture. Matching the $181\times181$ planar aperture would require $M=390$ antennas, yielding an MMPSR of $18.53$~dB, about $4$~dB higher than the proposed scheme but with roughly $10\times$ more antennas.

Next, we evaluate the antenna deactivation framework in Algorithm~\ref{alg:antenna_deactivation} for a linear array with $N=256$ locations. We consider far-field sensing, where the target ranges exceed the Fraunhofer distance and the array response is determined by target direction. Two targets are located at $\pm25^\circ$ within a sensing region of $[-60^\circ,60^\circ]$, sampled every $0.5^\circ$, with a $2^\circ$ guard interval around each target excluded from the sidelobe region. Algorithm~\ref{alg:SA_transmit} first selects $M=16$ active locations, from which the MMPSR lookup matrix $\boldsymbol{\Gamma}$ and backtracking matrix $\mathbf{B}$ are constructed using Algorithm~\ref{alg:antenna_deactivation}.
Fig.~\ref{fig:min_active_ants} shows the minimum number of active antennas required by the proposed scheme, centered ULA, large-aperture ULA, and nested array in \cite{Pal2010nested} to meet different MMPSR requirements, with the TCM reoptimized for each tested number of active antennas. The proposed scheme requires substantially fewer antennas; for example, achieving $14$~dB requires $14$ antennas, compared with $38$, $46$, and $73$ for the centered ULA, nested array, and large-aperture ULA, respectively. The centered ULA also requires fewer antennas than the latter two because, in the far field, its uniform $\lambda/2$ spacing avoids the angular ambiguities associated with their larger inter-element spacings.

\vspace{-2mm}
 \section{Conclusions}
In this paper, we proposed an energy-efficient non-uniform array design framework for multi-target sensing. The proposed framework jointly optimizes the antenna configuration and TCM to maximize the MMPSR. Furthermore, an adaptive antenna deactivation strategy was developed to switch off additional antennas while meeting a given MMPSR requirement, thereby providing a flexible trade-off between sensing performance and energy consumption.
Numerical results demonstrated the superior performance of the optimized non-uniform array over uniform configurations for the same number of active antennas. 
Furthermore, the proposed SA-based non-uniform array design, together with the deactivation strategy, significantly reduces the number of active antennas required to achieve a required sensing performance.

\vspace{-2mm}
\bibliographystyle{IEEEtran}
\bibliography{refs} 

@IEEEtranBSTCTL{BSTcontrol,
  CTLuse_forced_etal       = "yes",
  CTLmax_names_forced_etal = "3",   
  CTLnames_show_etal       = "1"   
}

@ARTICLE{Pal2010nested,
  author={Pal, Piya and Vaidyanathan, P. P.},
  journal={IEEE Trans. Signal Process.}, 
  title={Nested Arrays: A Novel Approach to Array Processing With Enhanced Degrees of Freedom}, 
  year={2010},
  volume={58},
  number={8},
  pages={4167-4181},
  doi={10.1109/TSP.2010.2049264}}

@ARTICLE{Pal2012nested,
  author={Pal, Piya and Vaidyanathan, P. P.},
  journal={IEEE Trans. Signal Process.}, 
  title={Nested Arrays in Two Dimensions, Part {I}: Geometrical Considerations}, 
  year={2012},
  volume={60},
  number={9},
  pages={4694-4705},
  doi={10.1109/TSP.2012.2203814}}

@ARTICLE{Rutenbar1989Simulated,
  author={Rutenbar, R.A.},
  journal={IEEE Circuits Devices Mag.}, 
  title={Simulated annealing algorithms: an overview}, 
  year={1989},
  volume={5},
  number={1},
  pages={19-26},
  doi={10.1109/101.17235}}

@ARTICLE{Yang2024Signal,
  author={Yang, Jiyuan and Chen, Yan and Gao, Xiqi and T. M. Slock, Dirk and Xia, Xiang-Gen},
  journal={IEEE Trans. Signal Process.}, 
  title={Signal Detection for Ultra-Massive {MIMO}: An Information Geometry Approach}, 
  year={2024},
  volume={72},
  number={},
  pages={824-838},
  doi={10.1109/TSP.2024.3358617}}

@ARTICLE{Lu2022NearField,
  author={Lu, Haiquan and Zeng, Yong},
  journal={IEEE Commun. Lett.}, 
  title={Near-Field Modeling and Performance Analysis for Multi-User Extremely Large-Scale {MIMO} Communication}, 
  year={2022},
  volume={26},
  number={2},
  pages={277-281},
  doi={10.1109/LCOMM.2021.3129317}}

@ARTICLE{Zhang2025Successive,
  author={Zhang, Zijian and Zhu, Jieao and Dai, Linglong and Heath, Robert W.},
  journal={IEEE Trans. Wireless Commun. }, 
  title={Successive {Bayesian} Reconstructor for Channel Estimation in Fluid Antenna Systems}, 
  year={2025},
  volume={24},
  number={3},
  pages={1992-2006},
  doi={10.1109/TWC.2024.3515135}}

@ARTICLE{Han2025Cell,
  author={Han, Tianyu and Zhu, Yongxu and Wong, Kai-Kit and Zheng, Gan and Shin, Hyundong},
  journal={IEEE Trans. Wireless Commun. }, 
  title={Cell-Free Fluid Antenna Multiple Access Networks}, 
  year={2025},
  volume={24},
  number={9},
  pages={7237-7251},
  doi={10.1109/TWC.2025.3559441}}

@ARTICLE{Zhu2023Movable,
  author={Zhu, Lipeng and Ma, Wenyan and Zhang, Rui},
  journal={IEEE Commun. Lett.}, 
  title={Movable-Antenna Array Enhanced Beamforming: Achieving Full Array Gain With Null Steering}, 
  year={2023},
  volume={27},
  number={12},
  pages={3340-3344},
  doi={10.1109/LCOMM.2023.3323656}}

@ARTICLE{Murino1996Synthesis,
  author={Murino, V. and Trucco, A. and Regazzoni, C.S.},
  journal={IEEE Trans. Signal Process.}, 
  title={Synthesis of unequally spaced arrays by simulated annealing}, 
  year={1996},
  volume={44},
  number={1},
  pages={119-122},
  doi={10.1109/78.482017}}

@article{Skolnik1964Planar,
  title={Planar arrays with unequally spaced elements},
  author={Skolnik, MI and Sherman, JW},
  journal={Radio Electron. Eng.},
  volume={28},
  number={3},
  pages={173--184},
  year={1964},
  publisher={IET}
}

@ARTICLE{Ning2025Movable,
  author={Ning, Boyu and Yang, Songjie and Wu, Yafei and Wang, Peilan and Mei, Weidong and Yuen, Chau and Bj\"ornson, Emil},
  journal={IEEE Wireless Commun. }, 
  title={Movable Antenna-Enhanced Wireless Communications: General Architectures and Implementation Methods}, 
  year={2025},
  volume={32},
  number={5},
  pages={108-116},
  doi={10.1109/MWC.013.2400238}}

@ARTICLE{Li2026Low,
  author={Li, Danqi and Tuan, Hoang Duong and Yu, Hongwen and Shu, Feng and Zhu, Wei and Shin, Hyundong and Wong, Kai-Kit},
  journal={IEEE Trans. Wireless Commun. }, 
  title={Low-Complexity Path-Following Optimization for Fluid Antennas and Beamforming in Multi-User Communication}, 
  year={2026},
  volume={25},
  number={},
  pages={18640-18654},
  doi={10.1109/TWC.2026.3703130}}

@ARTICLE{Emil2025Enabling,
  author={Bj\"ornson, Emil and Kara, Ferdi and Kolomvakis, Nikolaos and Kosasih, Alva and Ramezani, Parisa and Salman, Murat Babek},
  journal={IEEE Open J. Commun. Soc. }, 
  title={Enabling 6{G} Performance in the Upper Mid-Band by Transitioning From Massive to Gigantic {MIMO}}, 
  year={2025},
  volume={6},
  number={},
  pages={5450-5463},
  doi={10.1109/OJCOMS.2025.3576931}}

@article{Wang2026Harnessing,
  title={Harnessing the Freedom of Non-Uniformity in Monostatic {ISAC} with Antenna Flexibility},
  author={Wang, Zhe and Zaher, Mahmoud and Petrov, Vitaly and Bj{\"o}rnson, Emil},
  journal={arXiv preprint arXiv:2604.27571},
  year={2026}
}

@ARTICLE{Palaiologos2024Robust,
  author={Palaiologos, Michail and Castañeda García, Mario H. and Kakkavas, Anastasios and Stirling-Gallacher, Richard A. and Caire, Giuseppe},
  journal={IEEE Trans. Wireless Commun. }, 
  title={Robust Non-Uniform {LoS MIMO} Array Design}, 
  year={2024},
  volume={23},
  number={10},
  pages={13202-13218},
  doi={10.1109/TWC.2024.3399452}}
\end{document}